\documentclass[prl,twocolumn,reprint,superscriptaddress,amsmath,floatfix]{revtex4-1}
\usepackage{graphicx}
\usepackage{amssymb}
\usepackage[percent]{overpic}
\begin{document}
\title{Cooperative enhancement of superconducting correlations by electron-electron
and electron-phonon interactions in the quarter-filled band}
\author{R. T. Clay}
\email{r.t.clay@msstate.edu}
\author{D. Roy}
\affiliation{Department of Physics \& Astronomy and HPC$^2$ Center for Computational Sciences,
  Mississippi State University, Mississippi State, MS 39762}
\date{today}
\vskip 1pc
\begin{abstract}
  We present the results of Quantum Monte Carlo calculations for a two
  dimensional frustrated Hubbard model coupled to bond phonons.  The
  model is known to have a $d$-wave superconducting ground state in
  the limit of large phonon frequency for sufficiently strong
  electron-phonon coupling.  In the absence of electron-phonon
  coupling the Hubbard interaction $U$ enhances superconducting
  pairing in the quarter-filled (density $\rho=0.5$) band.  We show
  here that at $\rho=0.5$ electron-electron and electron-phonon
  interactions cooperatively reinforce $d$-wave pairing, while
  competing with each other at all other densities.  Cooperative
  degrees of freedom are found in many phase transitions and are
  essential to understanding superconductivity in strongly correlated
  materials.
\end{abstract}
\maketitle

{\it Introduction.---}There is strong evidence that any successful
theory of superconductivity (SC) in the high T$_c$ cuprates and other
unconventional superconductors must incorporate the effects of
electron-electron (e-e) interactions. It is now often assumed that
superconducting pairing in these materials can be mediated solely by
e-e interactions, leading to numerous studies of model Hamiltonians
with e-e interactions such as the Hubbard model.  While the properties
of the two dimensional (2D) Hubbard model remain under active debate,
calculations with the best unbiased methods available have found no
long-range superconducting pairing in the weakly doped Hubbard model
on a square lattice \cite{Ehlers17a,Qin19a} or the half-filled Hubbard
model on the anisotropic triangular lattice \cite{Dayal12a}.

One possible reason that such calculations have failed to find SC is
that the superconducting state is of a more complex form than
conventionally assumed, for example a pair density wave
\cite{Anderson04b,Franz04a,Tesanovic04a,Chen04a,Vojta08a,Berg09a}. It
is also possible that repulsive e-e terms alone are not sufficient for
SC and that other interactions must be considered.  The most obvious
candidate is the electron-phonon (e-p) interaction.  In the cuprates
there is strong experimental evidence for the importance of phonons in
the electronic properties and SC.  This includes giant softening of
the Cu-O bond stretching frequency in the underdoped cuprates
\cite{Reznik10a}, the oxygen isotope effect
\cite{Hofer00a,Bendele17a}, and kinks observed in photoemission
experiments \cite{Lanzara01a}. Phonons also appear to play some role
in the ubiquitous charge ordering found across the cuprate family
\cite{Reznik10a,LeTacon13a}.

Away from the weakly doped half-filled band there have been fewer
investigations of SC in interacting models.  Calculations on
frustrated lattices have shown that e-e interactions enhance
superconducting pairing at quarter-filling (corresponding to a carrier
density of $\rho=0.5$) while suppressing pairing at all other
densities \cite{Gomes16a,DeSilva16a,Clay19b}.  Clay and collaborators
have proposed a valence-bond theory of SC at $\rho=0.5$
\cite{Clay19a}. In this picture SC emerges from a proximate
spin-gapped insulating state, similar to the proposal of resonating
valence-bond (RVB) SC emerging from a 2D valence-bond solid (VBS)
\cite{Anderson87a}.  The RVB theory of SC in the weakly doped Hubbard
model remains controversial, and as noted above calculations for
$\rho\lesssim1$ have not found SC.  In two dimensions
antiferromagnetism (AFM) is preferred over singlet formation and
calculations at $\rho=1$ have also failed to find a VBS state
\cite{Gomes13a}.

A valence bond insulator, the Paired Electron Crystal (PEC), does occur at
$\rho=0.5$ \cite{Li10a,Dayal11a,Clay19a}. The PEC is a density wave of
pairs, with nearest-neighbor (n.n.)  singlet pairs separated by pairs
of vacant sites.  The PEC is characterized by a spin gap and coexisting charge and
bond order, and becomes favored over  AFM at $\rho=0.5$
in the presence of lattice frustration \cite{Li10a,Dayal11a}.  The PEC
is often adjacent to SC in the organic charge-transfer solids (CTS),
which are all $\rho=0.5$ materials \cite{Clay19a}.  Besides the CTS
many other $\rho=0.5$ superconductors are known \cite{Clay19a}, and
this mechanism of SC has been further generalized to the cuprates
\cite{Mazumdar18a}.  E-p interactions are required to realize the PEC
because of the bond distortion that simultaneously occurs with
n.n. singlet formation at $\rho=0.5$ \cite{Li10a,Dayal11a}.  It is
 natural to expect that SC evolving from a destabilized PEC must
involve both e-e and e-p interactions.

{\it Model.---}In many theories e-p and e-e interactions have been
considered to be mutually exclusive in mediating SC.  This belief is
partly due to the failures of the phonon-based pairing mechanism of
the BCS model in unconventional superconductors \cite{Stewart17a}, but
is also reinforced by the study of Hamiltonians that combine e-e and
e-p interactions.  The simplest is the Hubbard-Holstein model (HHM),
where at each site a dispersionless phonon is coupled to the charge
density:
\begin{eqnarray}
  H &=& -\sum_{\langle i,j\rangle,\sigma}t_{ij}(c^\dagger_{i,\sigma}c_{j,\sigma}+H.c.) 
  + U\sum_i n_{i,\uparrow}n_{i,\downarrow}    \label{ham-hhm} \\
  &+& g\sum_{i,\sigma}x_in_{i,\sigma} 
   +\sum_i\left(\frac{p_i^2}{2M}+\frac{M\omega^2}{2}x_i^2\right). \nonumber
\end{eqnarray}
 In Eq.~\ref{ham-hhm} $c^\dagger_{i,\sigma}$ creates an electron of
 spin $\sigma$ on site $i$,
 $n_{i,\sigma}=c^\dagger_{i,\sigma}c_{i,\sigma}$, $t_{ij}$ is the
 electron hopping integral, and $U$ is the onsite Hubbard interaction.
 $x_i$ and $p_i$ are coordinate and momentum operators for the phonon
 oscillator at site $i$ with mass $M$ and frequency $\omega$.  The e-p
 coupling constant is $g$.  We give energies below in units of the
 bare hopping $t$.  The physics of the HHM is governed by the
 competition between onsite pairing driven by the e-p interaction and
 opposed by $U$.  This is most clearly seen in the limit
 $\omega\rightarrow\infty$ where the effective Hubbard interaction
 between electrons is $U_{\rm eff}=U-2g^2/\omega$.

 We consider here a model with dispersionless {\it bond}-coupled
  (Su-Schrieffer-Heeger-type \cite{Su79a}) phonons,
\begin{eqnarray}
  H & = & -\sum_{\langle i,j\rangle,\sigma}t_{ij}[1+\alpha x_{(ij)}](c^\dagger_{i,\sigma}c_{j,\sigma}+H.c.) \label{ham-ssh} \\
  & + & U\sum_i n_{i,\uparrow}n_{i,\downarrow} 
  +\sum_{\langle i,j\rangle}\left(\frac{p_{(ij)}^2}{2M} 
 +  \frac{M\omega^2}{2}x_{(ij)}^2\right). \nonumber 
\end{eqnarray}
In Eq.~\ref{ham-ssh} $x_{(ij)}$ is the phonon coordinate associated
with the bond connecting sites $i$ and $j$ and all the other terms
have identical meaning to Eq.~\ref{ham-hhm}.  Compared to the large
amount of work on the HHM there are few numerical studies of
bond-coupled phonons beyond the classical limit.  Most work at finite
$\omega$ has been in one dimension
\cite{Sengupta03a,Clay07a,Hohenadler16a,Hohenadler18a,Sous18a}; a 2D
multi-orbital lattice was studied recently, although only for $U=0$.
\cite{Li19a}.  The $\omega\rightarrow\infty$ limit of
Eq.~\ref{ham-ssh} on a square lattice results in a more complex
effective interaction than in the HHM, with four types of terms
\cite{Hirsch87a,Assaad96a,Sous18a}.  These include a nearest neighbor
repulsion for parallel spins, on-site pair terms (which will be
suppressed by $U$), and pair hopping of nearest-neighbor singlet
pairs. Recent work has shown that this pair hopping interaction leads
\begin{figure}
  \begin{center}
    \resizebox{3.3in}{!}{\includegraphics{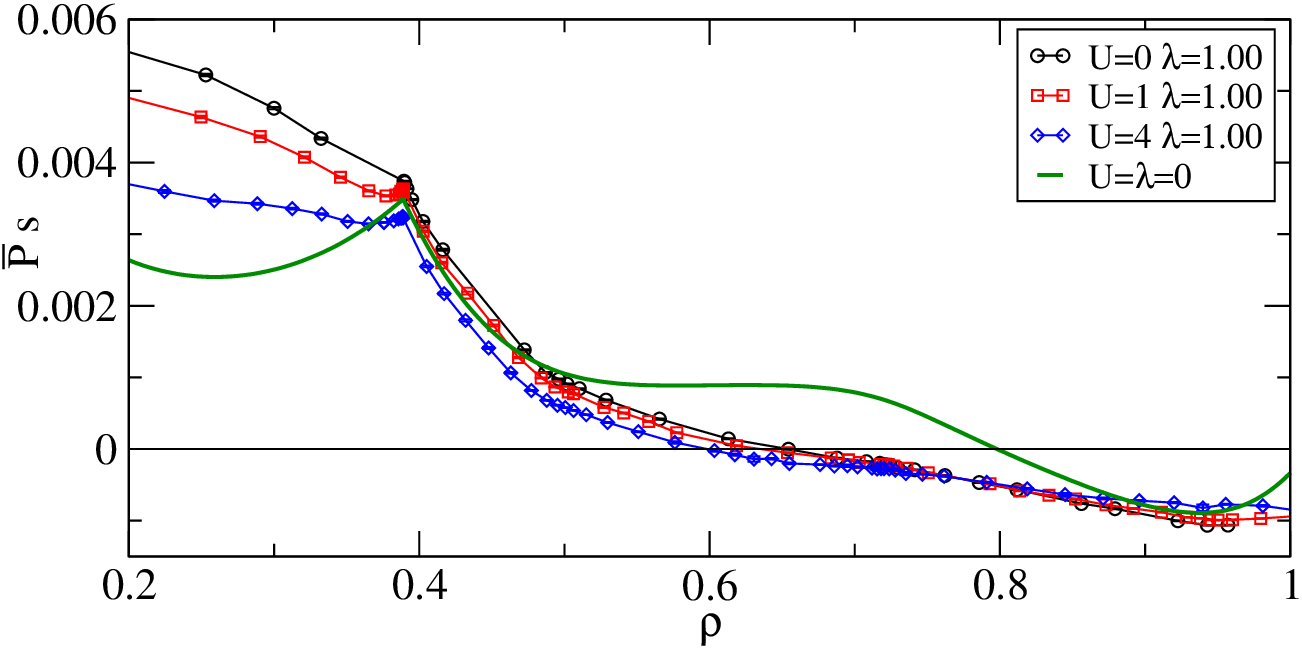}}
  \end{center}
  \caption{Average $s$ long-range pair-pair correlations
    $\bar{P}$ versus density $\rho$ for the 6$\times$6 anisotropic
    triangular lattice with $\omega=0.5$ at inverse temperature
    $\beta=8$.}
  \label{6x6rho-s}
\end{figure}  
to strongly bound bipolarons with light effective masses that are
stable against strong e-e interactions \cite{Sous18a}. This suggests
the possibility of e-p mediated SC in the presence of strong e-e
interactions at finite carrier densities.

The effective $\omega\rightarrow\infty$ Hamiltonian, $H_{\rm W}$, was
studied using quantum Monte Carlo near half-filling for $U=4$ on the
square lattice \cite{Assaad96a,Assaad97a,Assaad98a}.  In this density
region the pair hopping terms $d_{x^2-y^2}$ mediate pairing but
compete with the Mott insulating state driven by $U$
\cite{Assaad96a,Assaad97a,Assaad98a}.  While this shows that $d$-wave
SC can in principle result from e-p interactions, for $\rho\approx 1$
$d$-wave SC is only possible provided the coupling is strong enough to
overcome e-e interactions, again reinforcing the belief that e-e and
e-p interactions cannot simultaneously drive SC.  In this Letter we
investigate the combined effects of bond-phonon coupling and e-e
interactions on superconducting pair-pair correlations over a wide
density range including both $\rho=1$ and $\rho=0.5$. We find that at
$\rho=0.5$ e-e and e-p interactions {\it cooperatively} enhance
pairing.

{\it Numerical results.---}We define singlet pair creation operators $\Delta_i$,
\begin{equation}
  \Delta^\dagger_i = \sum_\nu \frac{g(\nu)}{\sqrt{2}}(c^\dagger_{i,\uparrow}c^\dagger_{i+\vec{r}_\nu,\downarrow}
  - c^\dagger_{i,\downarrow}c^\dagger_{i+\vec{r}_\nu,\uparrow}),
\end{equation}
where $g(\nu)$ is a relative sign that determines the pairing
symmetry.  The pair-pair correlation function is
$P(r)=\langle\Delta_i^\dagger\Delta_{i+\vec{r}}\rangle$. A theory of
correlated electron SC should satisfy two requirements, (i) at zero
temperature $P(r)$ must have long-range order, and (ii) $P(r)$ in the
presence of e-e interactions should be enhanced over its value for
non-interacting fermions. Finite and zero temperature calculations on
frustrated Hubbard models of up to 128 sites have shown that $U$
enhances $d$-wave pairing preferentially at $\rho=0.5$ while
suppressing pair-pair correlations relative to their $U=0$ value at
all other $\rho$ \cite{Gomes16a,DeSilva16a,Clay19b}.  While
enhanced by $U$ the magnitude of $P(r)$ however decreases with
distance, consistent with either a zero or possibly small long-range
superconducting order parameter \cite{Gomes16a,DeSilva16a,Clay19b}.

To solve Eq.~\ref{ham-ssh} we use finite-temperature determinant
quantum Monte Carlo (DQMC) which provides unbiased results
\cite{Gubernatis16a}.  The lowest temperatures that can be reached in
DQMC are strongly limited by the Fermion sign problem.  In the density
region of most interest here, $\rho\approx 0.5$, the sign problem is
considerably less severe than for the more intensively studied
$\rho\approx 0.8$.  Nevertheless, we are not able to reach low enough
temperatures to determine whether the ground state has long-range
superconducting order; however it is important to recall that the
effective model in the $\omega\rightarrow\infty$ limit {\it does} have
a superconducting ground state \cite{Assaad96a,Assaad97a,Assaad98a}.
Besides the sign problem, Monte Carlo autocorrelation times often
increase exponentially with e-p coupling near phase transitions
\cite{Hardikar07a}.  To help mitigate this we implemented the block
phonon updates of Reference \cite{Johnston13a}.  We use an imaginary
time discretization of $\Delta\tau=0.1$, which is small enough that
this source of systematic error can be neglected. We report results in
terms of the dimensionless e-p coupling strength
$\lambda=\alpha^2t_x/(M\omega^2)$ and set $M=1$.  Further information
on the method is given in the Supplemental information
\cite{Supplemental}.
\begin{figure}
  \begin{center}
    \resizebox{3.3in}{!}{\includegraphics{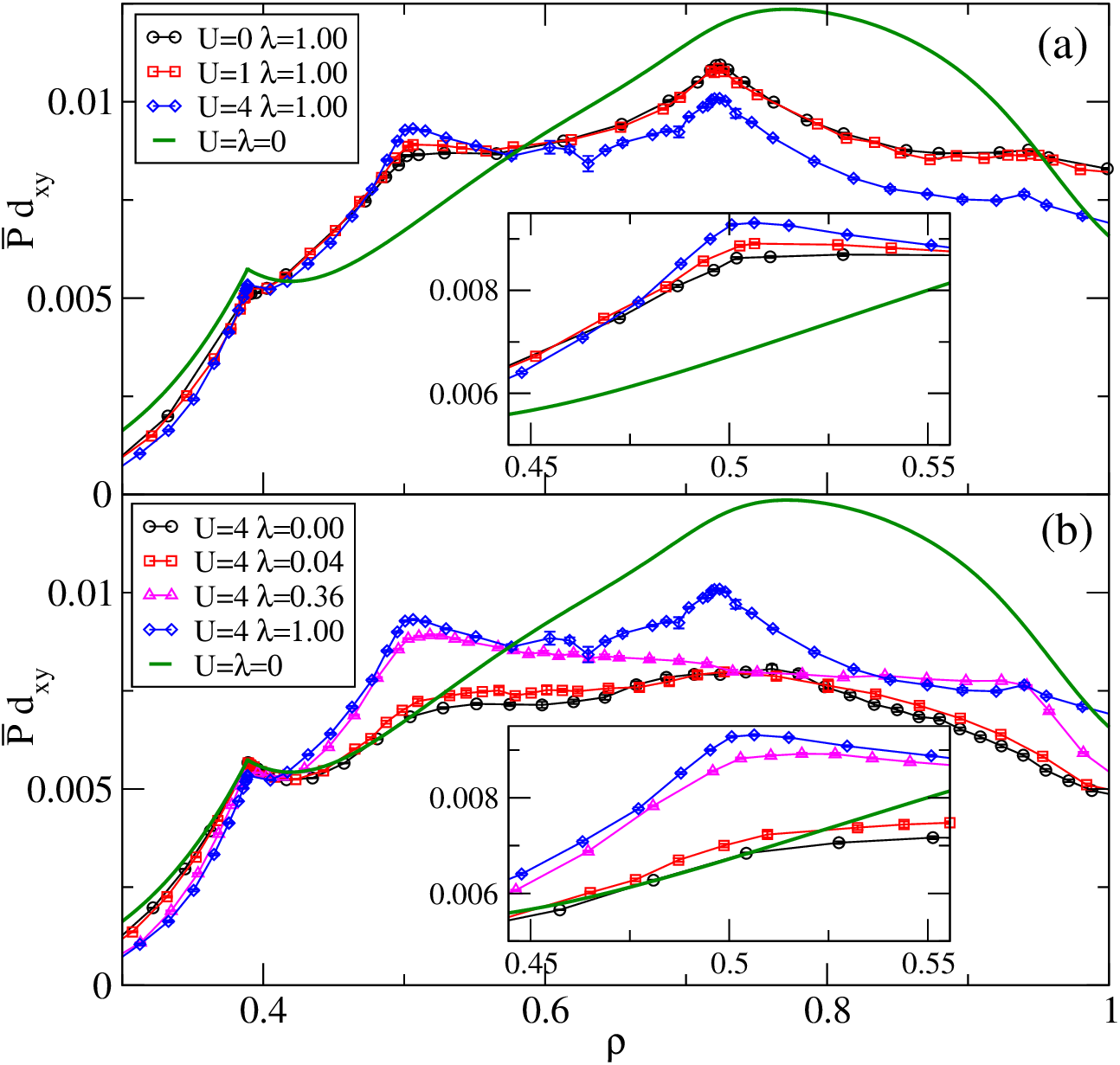}}
  \end{center}
  \caption{Same parameters as Fig.~\ref{6x6rho-s}, but for $d_{xy}$
    pairing.  (a) shows the effect of increasing $U$ at fixed
    $\lambda$.  (b) shows the effect of increasing $\lambda$ at fixed
    $U$.  Both interactions enhance $d_{xy}$ pairing at
    $\rho=0.5$. The insets magnify the density region around
    $\rho=0.5$.}
  \label{6x6rho}
\end{figure}  

We performed calculations on 4$\times$4, 6$\times$6, and 10$\times$10
anisotropic triangular lattices \cite{Gomes16a}.  This lattice has a
single frustrating bond $t^\prime$ across each plaquette; in the limit
$t^\prime=t$ ($t^\prime=0$) it is the triangular (square) lattice.
Because the PEC requires lattice frustration \cite{Li10a,Dayal11a}, we
work in the strongly frustrated limit with $t^\prime=0.8$. The lattice
dimensions are chosen under the constraint that the $\rho=0.5$
single-particle state is non-degenerate \cite{Gomes16a}. We also take
$t_y$ = 0.9 slightly different from $t_x$ = 1.0 in order to increase
the number of non-degenerate densities. This lessens the severity of
the Fermion sign problem and makes calculations feasible over a range
of $\rho$.  The precise choice of $t_x$, $t_y$, and $t^\prime$ are
however not critical to the results we report.  $P(r=0)$ can be
decomposed into combinations of charge and spin correlations
\cite{Aimi07a}.  AFM order leads to a trivial increase of the
short-range component of $P(r)$, even as the long-range component is
strongly suppressed \cite{Aimi07a,Dayal12a}.  To mitigate such
finite-size effects we exclude small $r$ correlations and measure the
average long-range value of $P(r)$ \cite{Huang01a},
\begin{equation}
  \bar{P}=N_p^{-1}\sum_{r>2}P(r).
  \label{pbar}
\end{equation}
In Eq.~\ref{pbar} only correlations over distances greater than two
lattice spacings are kept; $N_p$ is the number of such terms.  We
considered four pairing symmetries: $s$-wave involving n.n. pairs,
$s_{xy}$ with next-nearest-neighbor (n.n.n) pairs, $d_{x^2-y^2}$, and
$d_{xy}$.
\begin{figure}
  \begin{center}
    \resizebox{3.3in}{!}{\includegraphics{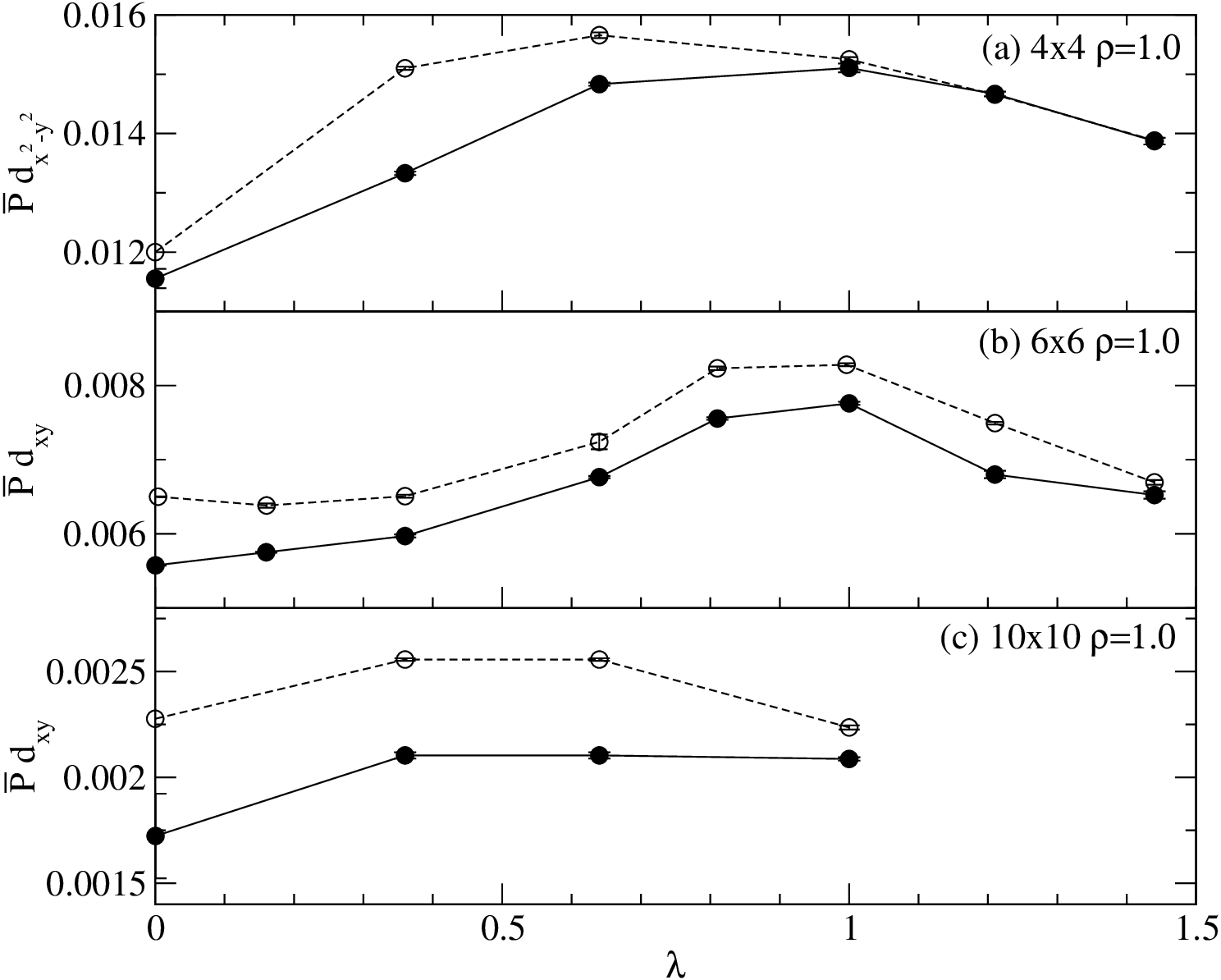}}
  \end{center}
  \caption{(a) $\bar{P}$ versus e-p coupling strength $\lambda$ for
    $\rho=1$, $\beta=8$, and $\omega=0.5$ on the (a) 4$\times$4, (b)
    6$\times$6, and (c) 10$\times$10 lattices.  Open (filled) symbols
    are for $U=0$ ($U=3$).}
  \label{rho1}
\end{figure}  

For the 4$\times$4 and 6$\times$6 lattices we calculated $\bar{P}$
over the density range 0.2 $\lesssim \rho \leq 1.0$.  In
Fig.~\ref{6x6rho-s} we plot $\bar{P}$ for $s$ pairing versus density
for the 6$\times$6 lattice (similar data for the 4$\times$4 lattice is
in the Supplemental information \cite{Supplemental}).  We find that at
all densities $U$ suppresses $s$ pairing; as $U$ increases $\bar{P}$
becomes closer to zero across the entire density range.  There is an
increase in $\bar{P}$ for $s$ pairing over its value for
non-interacting electrons for strong $\lambda$ in the low density
region $\rho\lesssim$ 0.4. An $s$ or $s_{xy}$ SC state may exist in
the model in the very low density region provided $U$ is not too
large; we will not consider this parameter region further here.
$s_{xy}$ pairing shows some enhancement by $\lambda$ for $\rho>0.4$,
but is also suppressed by $U$ \cite{Supplemental}.

In the thermodynamic limit on the anisotropic triangular lattice we
expect a pairing symmetry that mixes $d_{xy}$ and $d_{x^2-y^2}$; on
finite lattices either $d_{x^2-y^2}$ or $d_{xy}$ is favored
\cite{Gomes16a,Laubach15a}.  In Fig.~\ref{6x6rho} we plot $\bar{P}$
for the $d_{xy}$ symmetry versus density for the 6$\times$6
lattice. Fig.~\ref{6x6rho}(a) shows the effect of increasing $U$ at
fixed e-p coupling strength. Compared to the noninteracting system
(solid line), pairing correlations are enhanced by $U$ selectively at
$\rho=0.5$. At all other densities pairing is suppressed by
$U$. Fig.~\ref{6x6rho}(b) shows the effect of increasing e-p coupling
at fixed $U$. Here phonons enhance the pairing over a wide density
range for $\rho\gtrsim$ 0.4, including at both $\rho=0.5$ and
$\rho=1$. However, only at $\rho=0.5$ do e-e and e-p interactions 
  both enhance $\bar{P}$; at other densities the interactions
compete.
\begin{figure}
  \begin{center}
     \resizebox{3.3in}{!}{\includegraphics{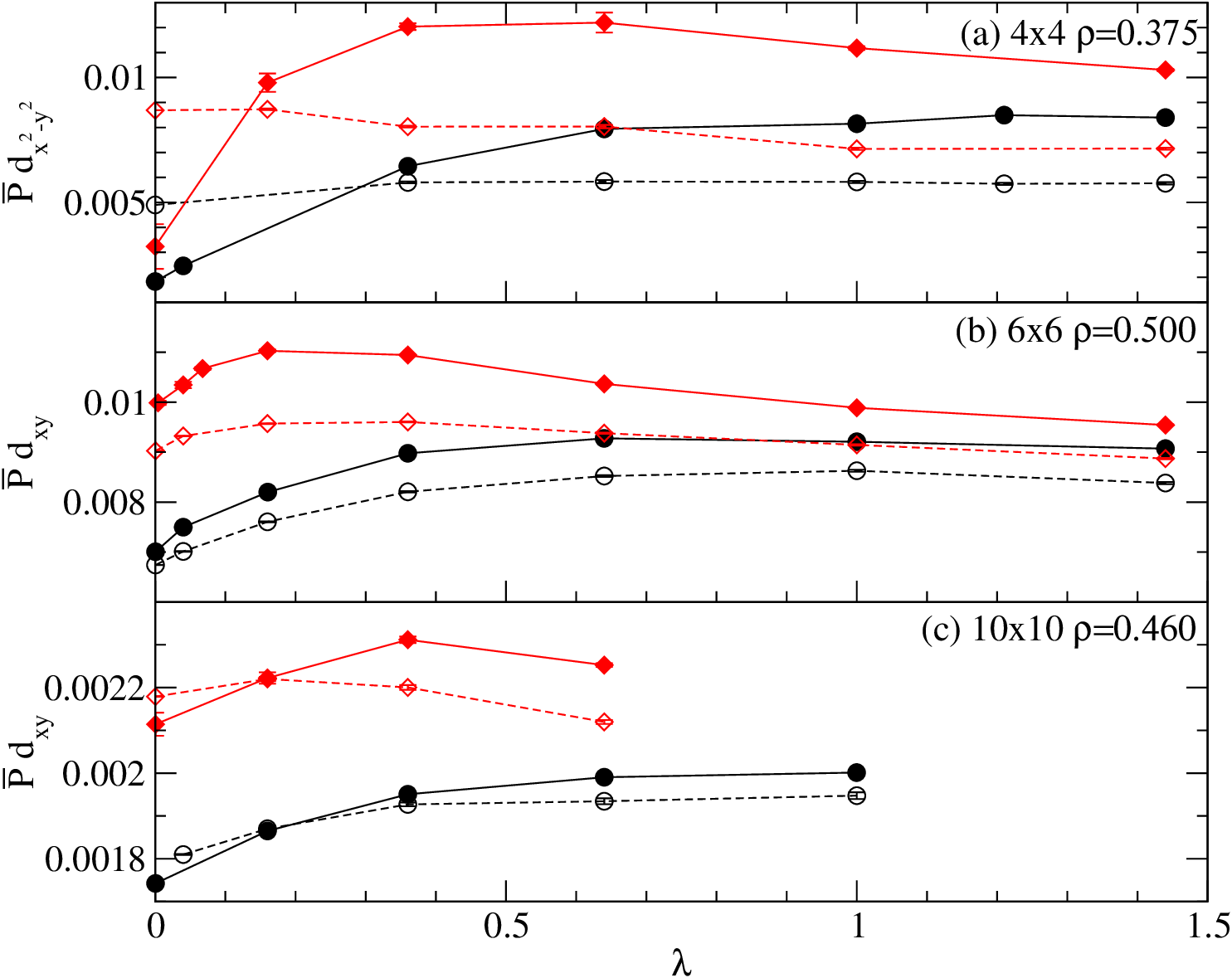}}
  \end{center}
    \caption{(a) $\bar{P}$ versus e-p coupling strength $\lambda$ for
      $\rho\sim0.5$ and $\omega=0.5$ on the (a) 4$\times$4, (b)
      6$\times$6, and (c) 10$\times$10 lattices.  Circles (diamonds)
      are for inverse temperature $\beta$ = 8 (16).  Open (filled)
      symbols are for $U=0$ ($U=3$).}
  \label{rho.5}
\end{figure}  

For the 10$\times$10 lattice we find that the Fermion sign is
reasonable for both $\rho\approx0.5$ and $\rho=1$.  Fig.~\ref{rho1}
summarizes our results for $\rho=1$ on all of the lattices.  At
$\rho=1$ $\bar{P}$ for either $d_{x^2-y^2}$ or $d_{xy}$ pairing
increases with $\lambda$ for $\lambda\lesssim 1$. This is consistent
with the $d$-wave superconducting state found in References
\cite{Assaad96a,Assaad97a,Assaad98a} in the $\omega\rightarrow\infty$
limit of the model. However, $U$ competes with the e-p interaction at
$\rho=1$ with $\bar{P}$ decreasing with increasing $U$. We expect on
less frustrated lattices at $\rho=1$ a competition between $d$-wave SC
mediated by the bond phonons and AFM order mediated by $U$.  On the
4$\times$4 lattice at large $\lambda$ $\bar{P}$ for $s_{xy}$ pairing
becomes comparable to $\bar{P}$ for $d_{x^2-y^2}$ pairing
\cite{Supplemental}. This may be the reason for the weak decrease of
the $d_{x^2-y^2}$ $\bar{P}$ with $U$ at large $\lambda$ in
Fig.~\ref{rho1}(a).

The behavior of $\bar{P}$ at $\rho=0.5$ (Fig.~\ref{rho.5}) is very
different from $\rho=1$. At $\rho\sim0.5$ zero-temperature
calculations find that $\bar{P}$ increases with $U$
\cite{Gomes16a,DeSilva16a,Clay19b}.  Here, for all the lattices we
find that $\bar{P}$ increases with increasing $U$ and $\lambda$ at the
{\it same} densities where $T=0$ calculations find enhancement by $U$
alone. Cooperative interactions however, should not merely both
increase the value of an order parameter, but the effect of the first
interaction should be strengthened in the presence of the second, and
vice versa.  This is indeed what we see at $\rho=0.5$: First, we see
that $\lambda$ enhances the effect of $U$: in Fig.~\ref{rho.5} the
increase in $\bar{P}$ between $U=0$ and $U=3$ for all lattices is
larger for $\lambda>0$ than at $\lambda=0$. In fact, at the
temperatures we can access here, on some lattices $\bar{P}$ decreases
with $U$ at $\lambda=0$.  Second, nonzero $U$ enhances the increase in
$\bar{P}$ with $\lambda$.  As a function of $\lambda$, $\bar{P}$
reaches a broad maximum at $\lambda=\lambda_{\rm max}$.  Comparing the
value of $\bar{P}$ at $\lambda=0$ and $\lambda_{\rm max}$, there is a
larger increase for $U>0$ compared to $U=0$.  These data show that at
$\rho\approx0.5$ $U$ and SSH phonon interactions not only both enhance
pairing, but their effect is {\it cooperative}.  This is the central
result of our work.

\begin{figure}
  \begin{center}
    \resizebox{3.3in}{!}{\includegraphics{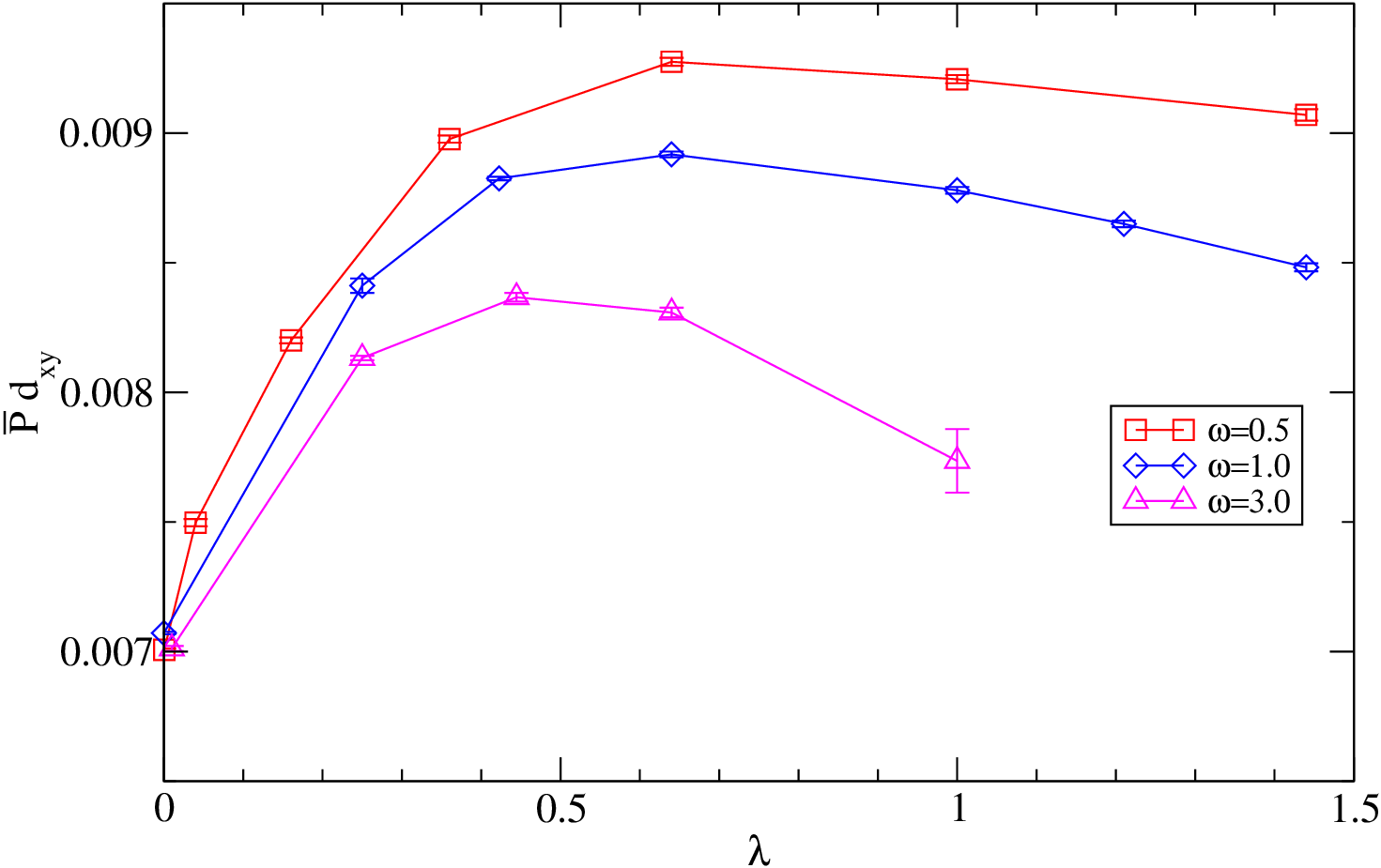}}
  \end{center}
  \caption{$\bar{P}$ versus e-p coupling strength $\lambda$ for
    $d_{xy}$ pairing in the 6$\times$6 lattice at $\rho=0.5$,
    $\beta=8$, and $U=3$. Squares, diamonds, and triangles correspond
    to phonon frequencies $\omega$ of 0.5, 1.0, and 3.0,
    respectively.}
  \label{6x6omega}
\end{figure}

In all lattices we find that as a function of $\lambda$, $\bar{P}$
reaches a broad peak at an intermediate value of $\lambda$ before
beginning to decrease.  A similar broad maximum in $\bar{P}$ is seen
as a function of $U$ in zero temperature $\lambda=0$ calculations at
$\rho=0.5$ \cite{DeSilva16a,Clay19b}.  We expect the decrease in
$\bar{P}$ at larger $\lambda$ and/or $U$ is caused by the increasing
effective mass of pairs.  $\lambda_{\rm max}$ and the amount of
enhancement also depend on $\omega$.  In Fig.~\ref{6x6omega} we show
the effect of the phonon frequency $\omega$ on $\bar{P}$ for the
6$\times$6 lattice. As $\omega$ decreases, $\lambda_{\rm max}$ shifts
towards stronger coupling and the amount of enhancement increases.
This shows that the phonon dispersion relation will play an important
role in understanding a superconducting state mediated by the
combination of e-e and e-p interactions.

{\it Conclusion.---}We have shown that at $\rho=0.5$ bond-coupled
phonons act cooperatively with onsite Coulomb interactions in
enhancing superconducting pairing.  Both interactions promote a
superconducting state through their effect on short-range (n.n.)
singlet pairing; the Hubbard $U$ through antiferromagnetic
superexchange, and bond-coupled phonons through effective
pair-hopping.  These interactions act cooperatively only for density
$\rho\approx 0.5$ while for all other densities they compete. The
presence of cooperative degrees of freedom is essential to understand
phase transitions in real materials, for example the
non-superconducting phases of the organic CTS \cite{Clay19a}.
Producing an unconventional superconducting state from e-p
interactions alone in the presence of competing e-e interactions would
require a careful tuning of parameters that is unrealistic.  As noted
above, model calculations have suggested that e-e interactions are not
sufficient for SC. Cooperative e-e and e-p interactions resolve both of these 
problems.

The observation of cooperative enhancement of pairing only at
$\rho\approx0.5$ supports our proposal of SC emerging from the PEC
\cite{Clay19a}.  The PEC has period four charge and bond order lattice
\cite{Li10a,Dayal11a}. This is only commensurate on the 4$\times$4
lattice, where with classical phonons a metal-PEC transition occurs at
a critical e-p coupling strength \cite{Li10a,Dayal11a}.  On the
4$\times$4 lattice we find the peak bond-bond correlation for
$\rho=0.5$ is at ${\bf Q}=(\pi/2,\pi)$, which is consistent with the
PEC \cite{Li10a,Dayal11a}. However, with the $\lambda$ and
temperatures accessible to DQMC we are not able to reach the metal-PEC
transition.  Inter-site Coulomb interactions $V_{ij}$ (below the
critical value $V_c$ for Wigner crystal formation) strengthen the PEC
and might be necessary to see the metal-PEC transition on large
lattices \cite{Dayal11a}.  Further calculations over a range of
$t^\prime$ and on commensurate lattices are in progress.  Studies of
the effective interaction in the $\omega\rightarrow\infty$ limit
across a wider density range will also be useful.

\begin{acknowledgments}
  We thank S. Mazumdar for useful discussions.  Some calculations in
  this work used the Extreme Science and Engineering Discovery
  Environment \cite{xsede} (XSEDE), which is supported by National
  Science Foundation grant number ACI-1548562. Specifically, we used
  the Bridges system \cite{bridges} which is supported by NSF award
  number ACI-1445606, at the Pittsburgh Supercomputing Center under
  awards TG-DMR190052 and TG-DMR190068. We also used the Comet system
  of the San Diego Supercomputer Center and
  the Stampede2 system of the Texas Advanced Computing Center under
  award TG-DMR190052.
\end{acknowledgments}

\end{document}